\documentclass{ws-ijmpe}

\begin{document}

\markboth{Jason Glyndwr Ulery For the STAR Collaboration}{Three-Particle Correlations from STAR}

\catchline{}{}{}{}{}

\title{THREE PARTICLE CORRELATIONS FROM STAR}

\author{\footnotesize Jason Glyndwr Ulery}

\address{Department of Physics, Purdue University, 525 Northwestern Avenue\\
West Lafayette, IN 47907,
USA\\
ulery@physics.purdue.edu}

\maketitle

\begin{history}
\received{(received date)}
\revised{(revised date)}
\end{history}

\begin{abstract}
Two-particle correlations have shown modification to the away-side shape in central Au+Au collisions relative to $pp$, d+Au and peripheral Au+Au collisions.   Different scenarios can explain this modification including:  large angle gluon radiation, jets deflected by transverse flow, path length dependent energy loss, Cerenkov gluon radiation of fast moving particles, and conical flow generated by hydrodynamic Mach-cone shock-waves.  Three-particle correlations have the power to distinguish the scenarios with conical emission, conical flow and Cerenkov radiation, from other scenarios.  In addition, the dependence of the observed shapes on the $p_T$ of the associated particles can be used to distinguish conical emission from a sonic boom (Mach-cone) and from QCD-\v{C}erenkov radiation. We present results from STAR on 3-particle azimuthal correlations for a high $p_T$ trigger particle with two softer particles.  Results are shown for $pp$, d+Au and high statistics Au+Au collisions at $\sqrt{s_{NN}}$=200 GeV.   An important aspect of the analysis is the subtraction of combinatorial backgrounds.  Systematic uncertainties due to this subtraction and the flow harmonics v2 and v4 are investigated in detail.  The implications of the results for the presence or absence of conical flow from Mach-cones are discussed.

\end{abstract}

\section{Introduction}
Though relativistic heavy ion collisions a medium is created that may be the quark gluon plasma (QGP).  We can study this medium though the use of jets and jet-like correlations.  Jets make good probes because their properties in vacuum can be calculated with perturbative quantum chromodynamics (pQCD).  Previous results on two-particle azimuthal jet-like correlations have revealed a broadened away-side shape in central Au+Au collisions relative to {\it pp}, d+Au and  peripheral Au+Au collisions, or even double humped \cite{2part1,2part2,2part3,2part4}.  The away-side shape is consistent with many different physics mechanisms including:  large angle gluon radiation \cite{glue1,glue2}, jets deflected by radial flow or preferential selection of particles due to path-length dependent energy loss, hydrodynamic conical flow generated by Mach-cone shock waves \cite{mach1,mach2}, and \v{C}erenkov gluon radiation \cite{cerenkov1,cerenkov2}.  3-particle correlations can be used to differentiate mechanisms with conical emission, Mach-cone and $\hat{C}$erenkov radiation, from the other mechanisms.   Additionally, the dependence of the conical emission angle on associated particle $p_T$ can be used to differentiate between Mach-cone and QCD-\v{C}erenkov radiation. 

\section{Analysis Procedure}

The 3-particle correlation analysis method has been rigorously described in reference \cite{proc}.  Results are reported here for trigger particles of $3<p_T<4$ Gev/c and associated particles of $1<p_T<2$ GeV/c, except where otherwise noted.  Results are from {\it pp}, d+Au, and Au+Au collisions at $\sqrt{s_{NN}}=200$ GeV/c.  All particles are charged particles measuremented in the STAR time projection chamber (TPC).   

\begin{figure}[htbp]
	\centering
		\includegraphics[width=1.0\textwidth]{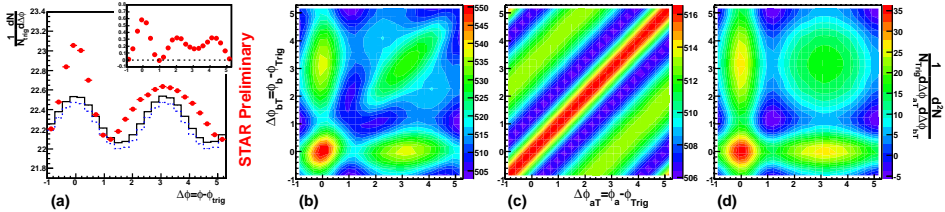}
			\vspace*{-0.0cm}
	\caption{(color online) (a) Raw 2-particle correlation (points), background from mixed events with flow modulation added-in (solid) and scaled by ZYA1 (dashed), and background subtracted 2-particle correlation (insert).  (b) Raw 3-particle correlation, (c) soft-soft background, $\beta \alpha^{2}B_{3}^{inc}$ and (d) hard-soft background + trigger flow, $\hat{J}_{2} \otimes \alpha B_{2}^{inc}$ + $\beta \alpha^{2} B_{3}^{inc,TF}$.  See text for detail.  Results are from ZDC-triggered 0-12\% Au+Au collisions at $\sqrt{s_{NN}}$=200 GeV/c.}
	\label{fig:Fig1}
\end{figure}

Figure~\ref{fig:Fig1}a shows the 2-particle azimuthal distribution ($J_2$), its background ($B_{2}^{inc}$) and the background subtracted 2-particle correlations ($\hat{J}_2$).  The background is constructed from mixed events where the trigger particle and the associated particles are from different events within the same centrality window.  The flow modulation is added in pairwise using the average $v_2$ values from the measurements based on the reaction plane and 4-particle cumulant methods \cite{2part1} and the $v_4$ contribution uses the parameterization $v_{4}=1.15v_{2}^{2}$ from the data\cite{reactionplane}.  The background is normalized (with scale factor $\alpha$) to the signal within $0.8<|\Delta\phi|<1.2$ (zero yield at 1 radian or ZYA1).

Figure~\ref{fig:Fig1}b shows the 3-particle azimuthal distribution ($J_3$) in $\Delta\phi_{aT}=\phi_{a}-\phi_{Trigger}$ and $\Delta\phi_{bT}=\phi_{b}-\phi_{T}$ where $\phi_{T}$, $\phi_a$, and $\phi_b$ are the azimuthal angles of the trigger particle and the two associated particles respectively.  Combinatorial backgrounds must be removed to extract the genuine jet-like 3-particle signal.  Events are treated as composed of two components, particles that are jet-like correlated with the trigger particle and background particles.  One source of background, the hard-soft background, results when one of the associated particles is jet-like correlated with the trigger particle and the other uncorrelated, other than the correlation due to flow.  It is constructed from the 2-particle jet-like correlations, $\hat{J}_2$ folded with the normalized 2-particle background, $\alpha B_{2}^{inc}$.  We shall refer to the hard-soft background as $\hat{J}_{2} \otimes \alpha B_{2}^{inc}$.

Another source of background, the soft-soft background, results from correlations between the two associated particles which are independent of the trigger particle.  This background is obtained from mixing the trigger particle with a different event of the same centrality.  We shall refer to this background as $B_{3}^{inc}$.  Since the two associated particle are from the same event all correlations between them that are independent of the trigger are preserved.  This may include contribution from minijets, other jets in the event and flow.  The soft-soft background is shown in figure~\ref{fig:Fig1}c.

Although the flow correlations between the two associated particles is accounted for by the soft-soft term, those between the associated particles and the trigger particle are not.  Those correlations are added in triplet-wise from mixed events where the trigger and the associated particles are all from different events in the same centrality window.  The $v_2$ and $v_4$ values are obtained from the same measurements as used in the 2-particle background.  The total number of triplets is determined from the soft-soft.  We shall refer to this background as $B_{3}^{inc,TF}$.

The total background is then, $\hat{J}_{2}\otimes \alpha B_{2}^{inc}$ + $\beta \alpha^{2}(B_{3}^{inc}+B_{3}^{inc,TF})$.  $B_{3}^{inc}$ and $B_{3}^{inc,tf}$ are scaled by $\beta \alpha^2$.  The normalization factor $\alpha$ is determined from 2-particle correlations and deviates from unity due to the combined effects of trigger bias and centrality definition.  If the events are poisson then $\alpha^2$ is the correct multiplicity scaling in 3-particle correlations.  The normalization factor $\beta$ corrects for the effect of non-poisson multiplicity distributions and is obtained such that the number of triplets in the background subtracted jet-like three-particles correlation equals the square of the number of pairs in the background subtracted jet-like 2-particle correlation.  Figure~\ref{fig:Fig1}d shows $\hat{J}_{2}\otimes \alpha B_{2}^{inc}$ + $\beta \alpha^{2}B_{3}^{inc,TF}$.

\section{Results}

\begin{figure}[htbp]
	\centering
		\includegraphics[width=1.0\textwidth]{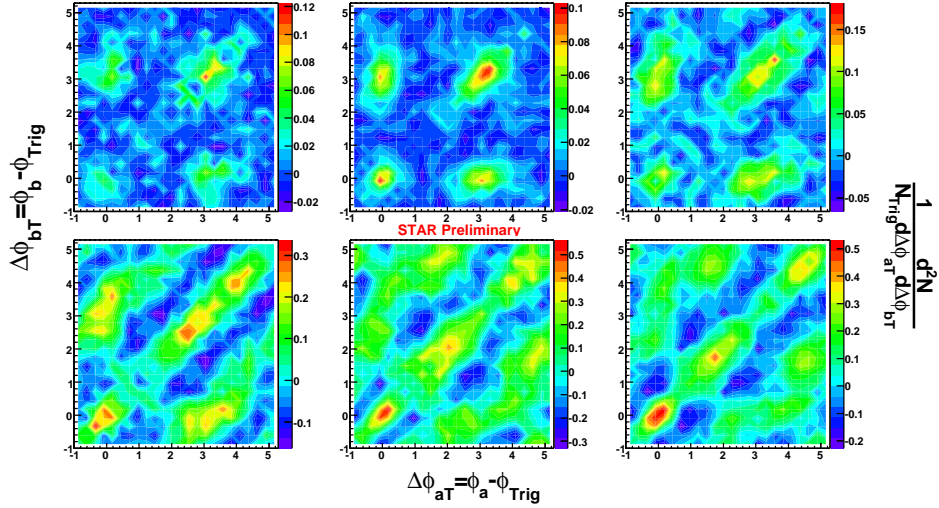}
			\vspace*{-0.0cm}
	\caption{(color online) Background subtracted jet-like 3-particle correlations for {\it pp} (top left), d+Au (top middle), and Au+Au 50-80{\%} (top right), 30-50{\%} (bottom left), 10-30{\%} (bottom center), and ZDC triggered 0-12{\%} (bottom right) collisions at $\sqrt{s_{NN}}$=200 GeV/c.}
	\label{fig:Fig2}
\end{figure}

Figure~\ref{fig:Fig2} shows background subtracted 3-particle jet-like correlation signals for different collisions and centralities.  The {\it pp} and d+Au results are similar with peaks clearly visible for the near-side, (0,0), away-side, ($\pi$,$\pi$), and the two cases of one particle on the near-side and the other on the away-side, (0,$\pi$) and ($\pi$,0).  The away-side peak displays on-diagonal elongation which is consistent with $k_T$ broadening.  Additional on-diagonal elongation is present in the Au+Au results, possibly due to deflected jets or large angle gluon radiation.  The more central Au+Au collisions display an off-diagonal structure, at about $\pi\pm1.45$ radians, that is consistent with conical emission.  This structure increases in magnitude with centrality and is prominent in the high statistics top 12\% data provided by the online zero degree calorimeter (ZDC) trigger.

\begin{figure}[htbp]
	\centering
		\includegraphics[width=1\textwidth]{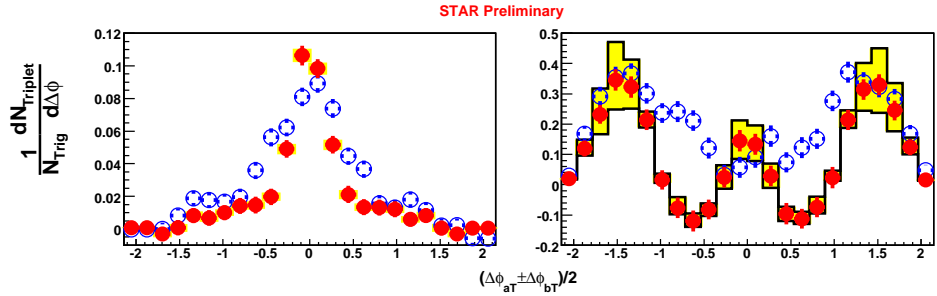}
			\vspace*{-0.0cm}
	\caption{(color online) Away-side projections of a strip of width 0.7 radians for (left) d+Au and (right) 0-12\% ZDC Triggered Au+Au.  Off-diagonal projection (solid) is $(\Delta\phi_{aT}-\Delta\phi_{bT})/2$ and on-diagonal projection (open) is $(\Delta\phi_{aT}+\Delta\phi_{bT})/2-\pi$. Shaded bands are systematic errors.}
	\label{fig:Fig5}
\end{figure}

Figure~\ref{fig:Fig5} shows away-side projections of on-diagonal strips to $(\Delta \phi_{aT} + \Delta \phi_{bT})/2 - \pi$ and off-diagonal strips to $(\Delta \phi_{aT} - \Delta \phi_{bT})/2$.  In d+Au collisions only a strong central peak is present for both on-diagonal and off-diagonal projections.  The on-diagonal projection is broader than the off-diagonal projection, likely due to $k_T$ broadening.  In central Au+Au collisions strong peaks are seen in the off-diagonal projection, as expected for conical emission.  The on-diagonal projection is similar to the off-diagonal but with additional contribution between the peaks likely due to deflected jets and/or large angle gluon radiation.  The fitted angle of the side peaks in the off-diagonal projection is about 1.45 radians.

\begin{figure}[htbp]
	\centering
		\includegraphics[width=1.0\textwidth]{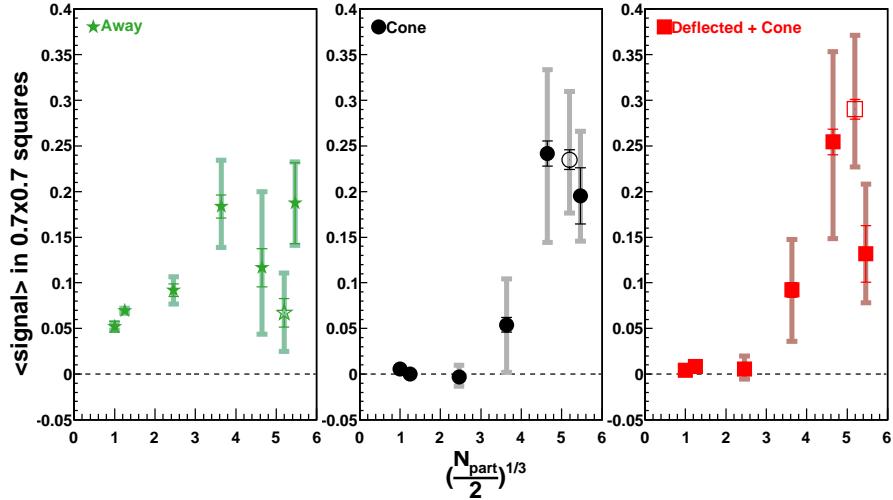}
			\vspace*{-0.0cm}
	\caption{(color online) Average signals in 0.7 $\times$ 0.7 boxes at (0,0), left, ($\pi\pm1.45$,$\pi\mp1.45$), center, and ($\pi\pm1.45$,$\pi\pm1.45$).  Solid error bars are statistical and shaded are systematic.  $N_{part}$ is the number of participants.  The ZDC 0-12\% points (open symbols) are shifted to the left for clarity.}
	\label{fig:Fig3}
\end{figure}

Figure~\ref{fig:Fig3} shows the centrality dependence of the average signal strengths in different regions.  The right panel shows the away-side signal, average singal centered at ($\pi$,$\pi$). It increases with centrality in {\it pp}, d+Au and perpherial Au+Au and then seems to level off for mid-central and central Au+Au collisions.   The middle panel shows the average signal where we only expect conical emission, at ($\pi\pm1.45$,$\pi\mp1.45$).  It increases with centrality and significantly deviates from zero in central Au+Au collisions.  The right panel shows the average signal were conical emissions deflected jets, and large angle gluon radiation could all contribute, at $\pi\pm1.45$,$\pi\pm1.45$).  This signal is similar to what we see where we only expect conical emission.  

\begin{figure}[htbp]
	\centering
		\includegraphics[width=0.6\textwidth]{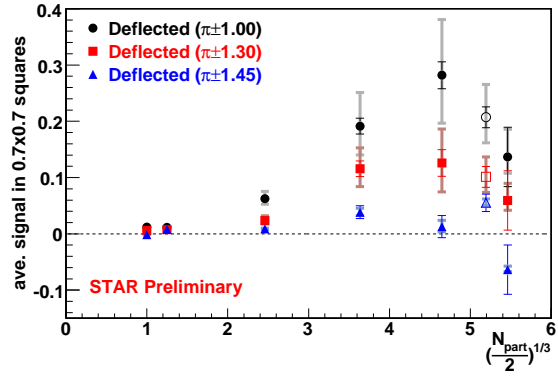}
			\vspace*{-0.0cm}
	\caption{(color online) Differences in average signals, between ($\pi\pm1.45$,$\pi\pm1.45$) and ($\pi\pm1.45$,$\pi\mp1.45$) (triangle), between ($\pi\pm1.3$,$\pi\pm1.3$) and ($\pi\pm1.3$,$\pi\mp1.3$) (square), and between ($\pi\pm1.0$,$\pi\pm1.0$) and ($\pi\pm1.0$,$\pi\mp1.0$) (circle).  Solid error bars are statistical and shaded are systematic.  $N_{part}$ is the number of participants.  The ZDC 0-12\% points (open symbols) are shifted to the left for clarity.}
	\label{fig:Fig4}
\end{figure}

Figure ~\ref{fig:Fig4} shows the difference between on-diagonal signals, where conical emission, deflected jets, and large angle gluon radiation could contribute, and off-diagonal signals, where only conical emission contributes.   Since conical emission signals are expected to be of equal magnitude on-diagonal and off-diagonal, the difference may indicate the contribution from deflected jets and large angle gluon radiation.  The centrality dependence is shown for three different angles.  The difference decreases with distance from ($\pi$,$\pi$).  

\begin{figure}[htbp]
	\centering
		\includegraphics[width=1\textwidth]{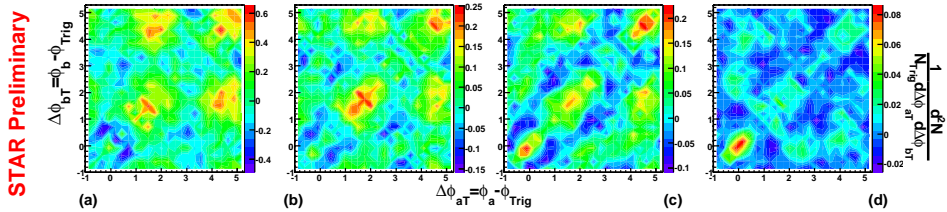}
			\vspace*{-0.0cm}
	\caption{(color online) Background subtracted jet-like 3-particle correlations for $0.75<p_{T}^{a,b}<1.0$ GeV/c (left), $1.0<p_{T}^{a,b}<1.5$ GeV/c (left center), $1.5<p_{T}^{a,b}<2.0$ GeV/c (right center), and $2.0 <p_{T}^{a,b}<3.0$ GeV/c (right).  Trigger particle $p_T$ is $3<p_T^T<4$ GeV/c.  Results are from ZDC triggered top 12\% central Au+Au collisons at $\sqrt{s_{NN}}$=200 GeV.}
	\label{fig:Fig6}
\end{figure}

\begin{figure}[htbp]
\hfill
\begin{minipage}[t]{.49\textwidth}
	\centering
		\includegraphics[width=1\textwidth]{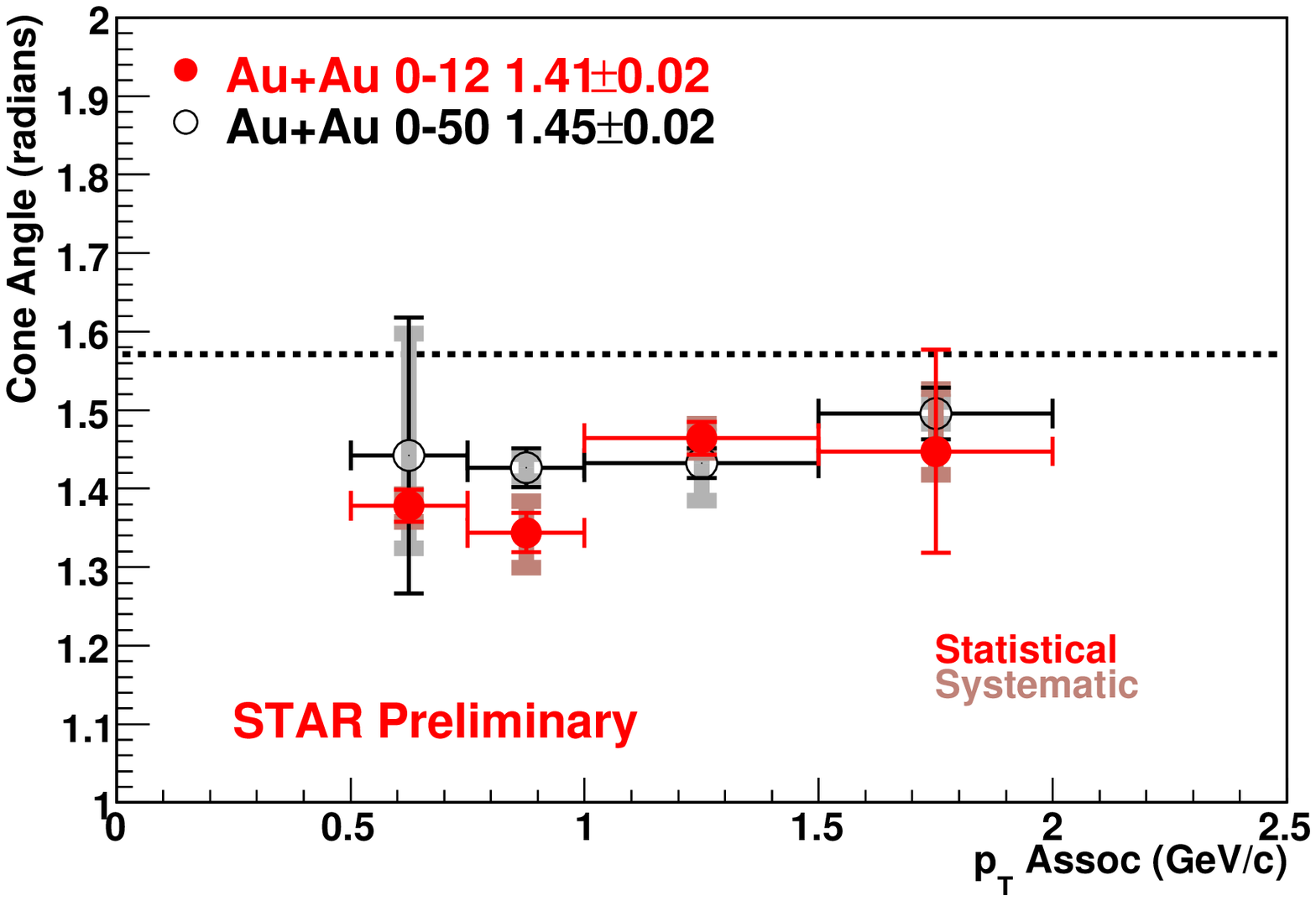}
			\end{minipage}
	\hfill
\begin{minipage}[t]{.49\textwidth}
	\includegraphics[width=1\textwidth]{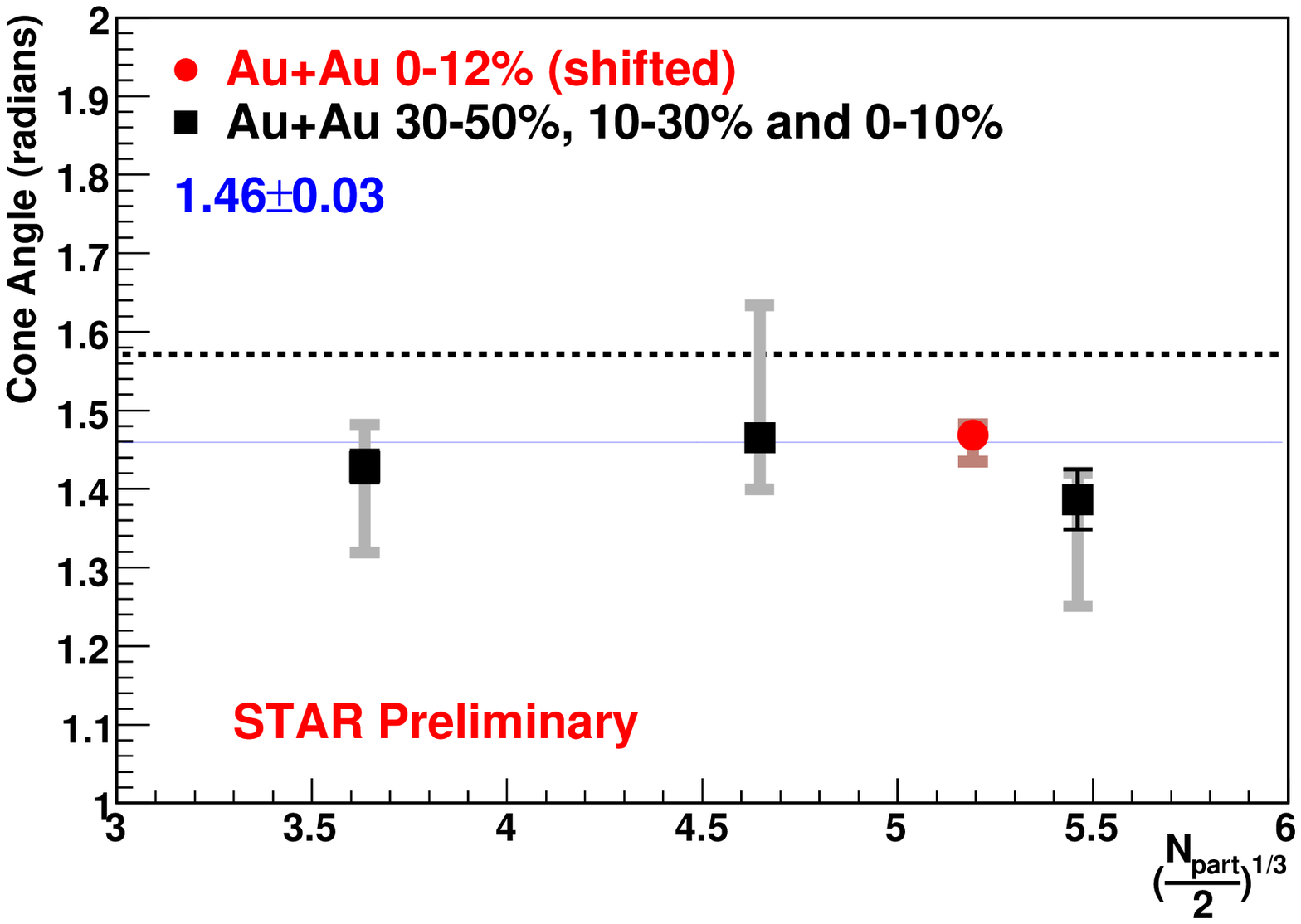}
			\end{minipage}
	\hfill
			\vspace*{-0.0cm}
	\caption{(color online) Emission angles from double Gaussian fits.  (left) Angle as a function of associated particle $p_T$ for Au+Au 0-12\% ZDC triggered (filled) and Au+Au 0-50\% from minimum bias (open).  Numbers on the plot are results from a fit to a constant for the data points with the fit errors displayed. (right) Angle as a function of centrality for Au+Au 0-12\% ZDC triggered data (circle) and Au+Au 30-50\%, 10-30\% and 0-10\% from minimum bias data (square).  The 0-12\% point has been shifted for clarity.  The number is from a fit to a constant for the points, shown with the solid line.  The dashed line is at $\pi/2$.  Solid error bars are statistical and shaded are systematic.}
	\label{fig:Fig7}	
\end{figure}

If we have Mach-cone emission, the emission angle is expected to be independent of the associated particle momentum; however, the \v{C}erenkov radiation model in Ref.~\cite{cerenkov2} predicts an emission angle that is sharply decreasing with associated particle momentum.  For this reason we shall look at the associated particle $p_T$ dependence of our signal.  Figure~\ref{fig:Fig6} shows the background subtracted 3-particle correlations for different associated $p_T$ bins. The angle is determined from fitting the off-diagonal projection, $(\Delta\phi_{aT}-\Delta\phi_{bT})/2$, to a central Gaussian and two symetric side Gaussians.  The strength of the off-diagonal signal decrease with increasing $p_T$ and is almost gone in the highest $p_T$ bin.  This is not surprising since we need two away-side particles each with a $p_T$ that is a significant fraction of the trigger particle $p_T$.   Figure~\ref{fig:Fig7} (left) shows the dependence of the angle of the off-diagonal peaks on associated particle $p_T$.  The angle is consistent with remaining constant as a function of associated particle $p_T$.

Figure~\ref{fig:Fig7} (right) shows the centrality dependence of the angle of the off-diagonal peaks obtained from fits to the projections as done for the $p_T$ dependence.  The angle is consistent with remaining constant from mid-central to central Au+Au collisions.  If we have Mach-cone emission this likely implies the speed of sound in the medium does not greatly vary from mid-central to central Au+Au collisions.  The solid line at 1.46 on the plot is from a fit to a constant.

\section{Systematics}

\begin{figure}[htbp]
	\centering
		\includegraphics[width=1\textwidth]{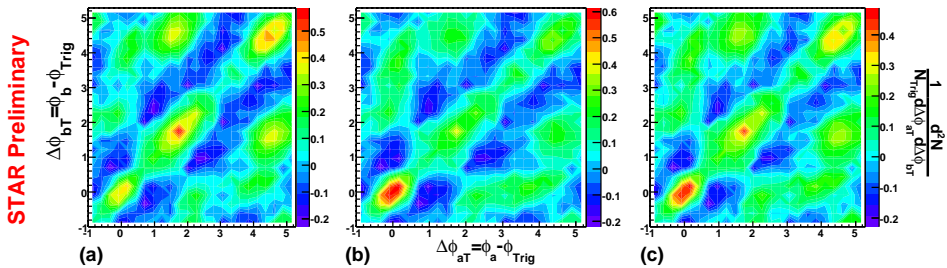}
			\vspace*{-0.0cm}
	\caption{(color online) 0-12\% Au+Au ZDC triggerd data for different systematic checks: (a) reaction plane $v_2$, (b) 4-particle cumulant $v_2$, and (c) normalization region for $\alpha$ of $0.6<|\Delta \phi|<1.4$}
	\label{fig:Fig8}
\end{figure}

The major sources of systematic error are from the elliptic flow measurements and the background normalization.  Our default $v_2$ is the average of measurements from the reaction plane and 4-particle cumulant methods.  The systematic uncertainty due to the $v_2$ has been determined by varying it between the reaction plane and 4-particle cumulant results.  Figure~\ref{fig:Fig8}a and b show the background subtracted 3-particle correlation for the reaction plane and 4-particle $v_2$, respectively.  Even though the hard-soft background and trigger flow backgrounds individually vary a great deal with the change in elliptic flow, the variations cancel out to first order.  Therefore the signal, as seen in Fig.~\ref{fig:Fig8} is robust with respect to the variation in elliptic flow.   

To study the effect of the background normalization the size of the normalization window used to determine $\alpha$ was doubled to $0.6<|\Delta \phi|<1.4$.  The signal is robust with this change in normalization.  Figure~\ref{fig:Fig8} shows the background subtracted 3-particle correlation using this larger normalization window.  

Other sources of systematic error include the effect on the trigger particle flow from requiring a correlated particle (a 20\% change on trigger particle $v_2$ is applied), uncertainity in the $v_4$ parameterization, and multiplicity bias effects on the soft-soft background.  The systematic errors shown in Figures~\ref{fig:Fig5},~\ref{fig:Fig3},~\ref{fig:Fig4}, and~\ref{fig:Fig7} reflect the quadratic sum of all the systematic uncertainties mentioned. 

\section{Conclusion}

Three-particle azimuthal correlations have been studied for trigger particles of $3<p_T<4$ GeV/c and associated particles of $1<p_T<2$ GeV/c in {\it pp}, d+Au, and Au+Au collisions at $\sqrt{s_{NN}}$=200 GeV/c by STAR.  This analysis treats events as the sum of two components, particles that are jet-like correlated with the trigger and background particles.  On-diagonal broadening has been observed in {\it pp} and d+Au collisions that is consistent with $k_T$ broadening.  Additional on-diagonal broadening has been observed in heavy ion collisions that may be due to contributions from deflected jets and/or large angle gluon radiation.  Off-diagonal peaks have been detected in central Au+Au collisions that are consistent with conical emission.  A study of the angle as a function of associated particle $p_T$ was performed to discriminate between hydrodynamic conical flow and \v{C}erenkov gluon radiation.  No strong dependence on associated particle $p_T$ was beheld.  This result is consistent with Mach-cone emission.

\end{document}